\documentclass[preprint,12pt]{elsarticle}

\usepackage{amsmath,amsfonts}
\usepackage{algorithmic}
\usepackage{array}
\usepackage[tight,footnotesize]{subfigure}
\usepackage{textcomp}
\usepackage{stfloats}
\usepackage{url}
\usepackage{verbatim}
\usepackage{graphicx}
\def\BibTeX{{\rm B\kern-.05em{\sc i\kern-.025em b}\kern-.08em
    T\kern-.1667em\lower.7ex\hbox{E}\kern-.125emX}}
\usepackage{balance}
\usepackage{booktabs}
\usepackage{caption}
\usepackage{makecell}
\usepackage{adjustbox}
\usepackage{wrapfig}
\usepackage[linesnumbered,ruled]{algorithm2e}
\usepackage{amssymb}
\usepackage{lineno}
\usepackage{float}
\usepackage{algorithmic}
\usepackage[tight,footnotesize]{subfigure}
\usepackage{booktabs}
\usepackage{caption}
\usepackage{makecell}
\usepackage{adjustbox}
\usepackage{wrapfig}
\usepackage{xcolor}
\usepackage[linesnumbered,ruled]{algorithm2e}

\journal{Journal Name}

\begin{document}
\sloppy
\setlength{\parskip}{0pt}

\begin{frontmatter}





\title{A Practical Guide to Agentic AI Transition in Organizations}

\author[label1]{Eranga Bandara}
\ead{cmedawer@odu.edu}

\author[label1]{Ross Gore}
\ead{rgore@odu.edu}

\author[label1]{Sachin Shetty}
\ead{sshetty@odu.edu}

\author[label7]{Sachini Rajapakse}
\ead{sachini.rajapakse@iciclelabs.ai}

\author[label7]{Isurunima Kularathna}
\ead{isurunima.kularathna@iciclelabs.ai}

\author[label10]{Pramoda Karunarathna}
\ead{pramodabhavani@sjp.ac.lk}

\author[label1]{Ravi Mukkamala}
\ead{mukka@odu.edu}

\author[label1]{Peter Foytik}
\ead{pfoytik@odu.edu}

\author[label1]{Safdar H. Bouk}
\ead{sbouk@odu.edu}

\author[label3]{Abdul Rahman}
\ead{abdulrahman@deloitte.com}

\author[label4]{Xueping Liang}
\ead{xuliang@fiu.edu}

\author[label8]{Amin Hass}
\ead{amin.hassanzadeh@accenture.com}

\author[label2]{Tharaka Hewa}
\ead{tharaka.hewa@oulu.fi}

\author[label5]{Ng Wee Keong}
\ead{awkng@ntu.edu.sg}

\author[label6]{Kasun De Zoysa}
\ead{kasun@ucsc.cmb.ac.lk}

\author[label9]{Aruna Withanage}
\ead{aruna@effectz.ai}
\author[label9]{Nilaan Loganathan}
\ead{nilaan@effectz.ai}

\address[label1]{Old Dominion University, Norfolk, VA, USA}
\address[label3]{Deloitte \& Touche LLP, USA}
\address[label4]{Florida International University, USA}
\address[label5]{Nanyang Technological University, Singapore}
\address[label6]{University of Colombo, Sri Lanka}
\address[label2]{Center for Wireless Communications, University of Oulu, Finland}
\address[label7]{IcicleLabs.AI}
\address[label10]{University of Sri Jayewardenepura, Sri Lanka}
\address[label8]{Accenture Technology Labs, Arlington, VA, USA}
\address[label9]{Effectz.AI}

\begin{abstract}

Agentic AI represents a significant shift in how intelligence is applied within organizations, moving beyond AI-assisted tools toward autonomous systems capable of reasoning, decision-making, and coordinated action across workflows. As these systems mature, they have the potential to automate a substantial share of manual organizational processes, fundamentally reshaping how work is designed, executed, and governed. Although many organizations have adopted AI to improve productivity, most implementations remain limited to isolated use cases and human-centered, tool-driven workflows. Despite increasing awareness of agentic AI's strategic importance, engineering teams and organizational leaders often lack clear guidance on how to operationalize it effectively. Key challenges include an overreliance on traditional software engineering practices, limited integration of business-domain knowledge, unclear ownership of AI-driven workflows, and the absence of sustainable human-AI collaboration models. Consequently, organizations struggle to move beyond experimentation, scale agentic systems, and align them with tangible business value. Drawing on practical experience in designing and deploying agentic AI workflows across multiple organizations and business domains, this paper proposes a pragmatic framework for transitioning organizational functions from manual processes to automated agentic AI systems. The framework emphasizes domain-driven use case identification, systematic delegation of tasks to AI agents, AI-assisted construction of agentic workflows, and small, AI-augmented teams working closely with business stakeholders. Central to the approach is a human-in-the-loop operating model in which individuals act as orchestrators of multiple AI agents, enabling scalable automation while maintaining oversight, adaptability, and organizational control.

\end{abstract}


\end{frontmatter}

\section{Introduction}

Recent advances in Large Language Models (LLMs) and AI tooling have accelerated the adoption of artificial intelligence across organizations~\cite{llm, llm-agents}. What began as experimentation with AI-assisted capabilities, such as code generation, content drafting, and decision support, has evolved toward systems capable of autonomous reasoning and coordinated action. These systems, commonly referred to as agentic AI, represent a fundamental shift in how intelligence is embedded within organizational workflows. Rather than merely assisting humans, agentic AI systems can act on their behalf within defined operational boundaries, enabling AI to participate directly in the execution of work~\cite{agentic-ai, agent-survey}.

As agentic AI systems mature, their impact extends beyond incremental productivity improvements toward the automation of entire manual and semi-manual processes. Activities that previously required continuous human coordination, such as information gathering, decision routing, operational monitoring, and exception handling, can increasingly be performed by networks of AI agents operating across heterogeneous systems. This shift has the potential to redefine how work is designed, how teams are organized, and how value is created within organizations~\cite{agentsway, agentic-workflow-practicle-guide}.

Despite this potential, most organizations remain in an intermediate transition phase rather than operating in a fully agentic state. Although AI adoption has become widespread, its use is often limited to isolated tools embedded within existing workflows~\cite{agentic-ai-rise}. In such settings, AI continues to function primarily as an assistant, with humans retaining responsibility for orchestration, decision-making, and execution. As a result, the transformative capabilities of agentic AI remain largely unrealized.

A central reason for this stagnation is that organizations approach agentic AI through the lens of traditional software development. Engineering teams often emphasize code-centric architectures, rigid interfaces, and deterministic logic, while business-domain knowledge, frequently encoded in informal processes, tacit expertise, and human judgment, remains weakly integrated into AI-driven systems. This disconnect leads to solutions that are technically functional but operationally misaligned with real organizational needs. Additional challenges further complicate the transition, including unclear ownership of AI-driven workflows, underdefined human--AI collaboration models, and growing concerns around trust, accountability, and control~\cite{towards-rai-xai}. At the same time, the rapid pace of AI research and tooling advancement creates a moving target, making it difficult for organizations to commit to long-term designs without fear of rapid obsolescence.

These challenges are not primarily technological. In practice, the dominant barriers to the adoption of agentic AI are organizational, cultural, and procedural. Successfully transitioning to agentic AI requires rethinking how work is structured, how teams are composed, and how responsibilities are distributed between humans and machines. Organizations must move beyond tool-centric adoption toward workflow-centric automation models that explicitly account for domain knowledge, human oversight, and continuous adaptation~\cite{llm-explainability, xai, responsible-ai}.

While our previous work~\cite{agentic-workflow-practicle-guide} focused on the engineering practices required to design, develop, and deploy production-grade agentic AI workflows, this paper shifts attention to the organizational and operational changes necessary for these systems to be effectively adopted and sustained in real-world settings. Drawing on practical experience building and deploying agentic AI workflows across multiple organizations and business domains, we present a pragmatic guide for transitioning from AI-assisted workflows to fully agentic AI systems.

The approach introduced in this paper emphasizes understanding business domains and manual processes, systematically delegating these processes to specialized AI agents, and keeping humans in the loop as orchestrators of agentic workflows. This transition is enabled through small, AI-augmented teams, deep collaboration between engineering and business stakeholders, and AI-assisted development practices in which AI systems themselves contribute to the construction and evolution of agentic workflows. By grounding agentic AI systems in human-centered operational models, organizations can achieve scalable automation while maintaining adaptability, accountability, and control. The main contributions of this work are summarized as follows.

\begin{enumerate}
    \item \textbf{Framing agentic AI adoption as an organizational transition problem rather than a purely technical challenge.}  
    This work characterizes the ongoing global transition toward agentic AI and explains why conventional, engineering-centric approaches are insufficient for realizing its full organizational impact.

    \item \textbf{Identification of key organizational and operational challenges in agentic AI transition.}  
    We analyze common obstacles observed in practice, including misalignment between engineering and business teams, insufficient integration of business-domain knowledge, unclear ownership of AI-driven workflows, and the absence of effective models for sustained human--AI collaboration.

    \item \textbf{A practical, experience-driven framework for transitioning to agentic AI systems.}  
    Drawing on real-world deployments across multiple organizations and business domains, we propose a pragmatic approach that emphasizes domain-driven use case identification, AI-assisted construction of agentic workflows, small AI-augmented development teams, close collaboration between business and engineering stakeholders, and the systematic translation of manual processes into human-supervised agentic workflows.

    \item \textbf{A human-centered operating model for scalable and sustainable agentic AI adoption.}  
    We introduce an operating model in which humans act as orchestrators of multiple AI agents, enabling scalable automation while preserving oversight, adaptability, and long-term organizational momentum.
\end{enumerate}

The remainder of this paper is organized as follows. Section~2 examines the key challenges organizations face during the transition to agentic AI, with an emphasis on organizational, procedural, and human-centered barriers that limit adoption beyond isolated AI-assisted use cases. Section~3 presents a practical guide for agentic AI transition, outlining principles and practices for identifying high-value use cases, delegating manual processes to AI agents, designing human-in-the-loop orchestration models, and enabling effective team structures and collaboration. Section~4 evaluates the proposed transition guide through a real-world tourism SME use case, assessing the effectiveness of agentic workflows in replacing manual operations and supporting human-supervised orchestration. Finally, Section~5 concludes the paper by summarizing key insights and contributions and outlining directions for future work, including the study of additional real-world agentic AI transition deployments across diverse organizational contexts.

\section{Challenges in Agentic AI Transition}

Organizations across industries are currently in an intermediate stage of AI adoption. While AI-assisted tools have become common in day-to-day work, most organizations have not yet transitioned to agentic AI systems capable of autonomously executing end-to-end workflows~\cite{agentic-ai-opptunities}. In practice, AI is predominantly used as a productivity enhancer for individuals rather than as an operational actor embedded within organizational processes.

A key reason for this stagnation is that many organizations have not fully realized the scope and implications of agentic AI. Agentic systems are capable of reasoning across multiple steps, interacting with heterogeneous systems, handling exceptions, and coordinating actions without continuous human intervention. These capabilities enable the automation of a substantial portion of existing manual and semi-manual workflows. However, this potential remains underappreciated, and agentic AI is often perceived as an incremental enhancement to existing tools rather than as a fundamental shift in how work is designed and executed~\cite{elderly-care-agentic-ai}. This misunderstanding gives rise to a set of recurring organizational and operational challenges, which are discussed in the following subsections.

\subsection{Limited Recognition of Agentic AI Capabilities}

Despite rapid advances in agentic AI technologies, many organizations have yet to fully recognize the breadth of tasks and workflows that can be automated using agentic systems. In practice, AI adoption is often confined to narrow use cases such as chat-based assistants, document summarization, or isolated task automation. While these applications provide incremental productivity benefits, they do not fundamentally change how work is executed within organizations~\cite{agentic-ai-opptunities}.

This limited recognition is partly driven by early exposure to large language models as conversational tools. As a result, AI is commonly perceived as an interface for information retrieval or text generation rather than as an autonomous actor capable of reasoning, planning, and executing complex workflows. Consequently, organizations underestimate the extent to which agentic AI can automate multi-step processes that involve coordination across systems, decision-making under uncertainty, and iterative refinement~\cite{agentic-ai, responsible-gen-ai}.

The impact of this constrained perception extends beyond technical design to organizational decision-making. Investment strategies, team structures, and success metrics are often aligned with low-risk, tool-centric deployments rather than transformative workflow automation. This cautious approach limits experimentation with agentic systems and discourages initiatives that could deliver substantial operational leverage.

Moreover, when agentic AI is framed primarily as an enhancement to individual productivity, its organizational value becomes difficult to quantify and justify at scale. End-to-end workflow automation, by contrast, can deliver measurable improvements in efficiency, consistency, and responsiveness. Failure to recognize this distinction leads organizations to underinvest in agentic AI and to overlook opportunities for meaningful transformation~\cite{agentic-ai-taxonomy-challenges}.

Addressing this challenge requires expanding the organizational understanding of what agentic AI can do. Leaders and practitioners must move beyond viewing AI as a collection of isolated tools and begin to conceptualize it as a workforce of autonomous agents capable of executing and coordinating complex workflows. Without this shift in perspective, organizations are unlikely to realize the full operational potential of agentic AI systems~\cite{xai-llm, towards-rai-xai}.

\subsection{Limited Understanding of Agentic AI and Related Concepts}

Despite rapid progress in agentic AI technologies, many development teams still have a limited and fragmented understanding of what agentic AI systems are and how they fundamentally differ from traditional LLM usage~\cite{llm, vision-language-model}. Because these concepts are relatively new, organizations often conflate agentic AI with simple prompt-based interactions or chatbot-style systems, significantly underestimating their capabilities and design implications~\cite{gpt-llm}.

\begin{figure}[H]
\centering
\includegraphics[width=5.3in]{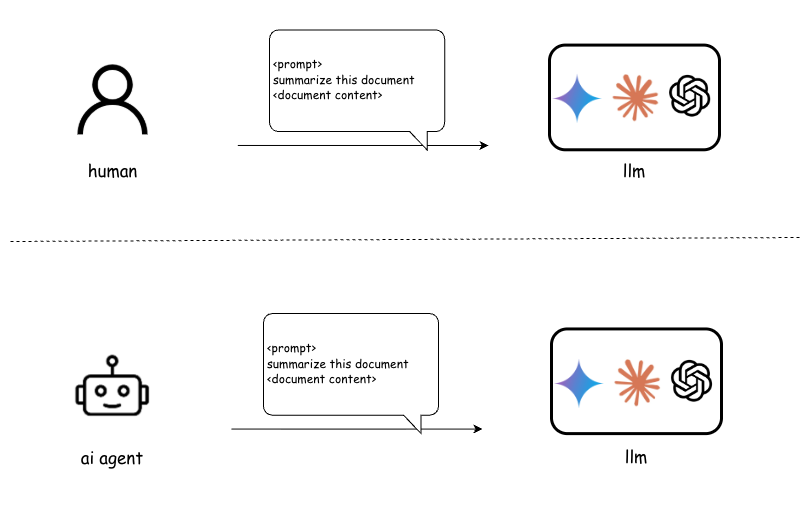}
\vspace{-0.1in}
\caption{Human--LLM interaction versus autonomous AI agent--LLM interaction.}
\label{ai-agent}
\end{figure}

Traditional LLM interactions follow a simple request--response pattern in which a human provides a prompt and the model generates a corresponding output, as illustrated in the top half of Figure~\ref{ai-agent}. In this paradigm, humans remain responsible for orchestration, decision-making, and follow-up actions. As a result, AI is treated primarily as a passive assistant rather than as an active participant in workflow execution~\cite{agentic-ai-workflow-patterns}.

In contrast, an AI agent can autonomously perform this interaction loop. An agent can construct prompts, invoke models, interpret responses, and trigger subsequent actions without direct human intervention, as illustrated in the bottom half of Figure~\ref{ai-agent}. In essence, AI agents are software programs that leverage LLMs in combination with tools, APIs, and external context to execute tasks automatically and iteratively~\cite{deep-psychiatric}.

When multiple such agents collaborate, each assigned specialized responsibilities such as searching, filtering, scraping, reasoning, validation, or publishing, they form agentic AI workflows~\cite{agentsway}. These workflows enable systems that can reason over complex tasks, plan sequences of actions, interact with external systems, monitor outcomes, and adapt their behavior through iterative feedback. Modern agentic workflows integrate LLMs with structured memory, search functions, databases, Model Context Protocol (MCP) servers, cloud services, and API-driven environments~\cite{deep-stride, mcp1, mcp2}. Rather than relying on a single monolithic prompt, responsibilities are distributed across specialized agents, improving modularity, controllability, and maintainability.

The limited understanding of these concepts leads organizations to design agentic systems using inappropriate abstractions, treating them as conventional software components or static AI services. This misunderstanding not only constrains system capabilities but also reinforces engineering-centric approaches that fail to exploit the full potential of agentic AI~\cite{nurolense}. Addressing this gap in conceptual understanding is a prerequisite for effective organizational transition to agentic AI workflows.

\subsection{Traditional Software Engineering Mindsets}

A significant barrier to effective agentic AI adoption is the tendency of engineering teams to approach agentic workflows using traditional software engineering paradigms. These paradigms emphasize rigid specifications, deterministic control flow, static interfaces, and extensive upfront design~\cite{agentic-ai-taxonomy-challenges}. While such approaches have proven effective for conventional software systems, they are poorly suited to agentic AI systems, which are inherently probabilistic, adaptive, and prompt-driven.

When agentic AI workflows are treated as traditional software components, teams often engage in overengineering, attempting to exhaustively define behavior, optimize prematurely, and impose strict structural constraints~\cite{ai-agent-hci}. This results in slow iteration cycles, brittle implementations, and systems that fail to exploit the flexibility and autonomy that agentic AI enables. Rather than simplifying work, these designs frequently reintroduce complexity in new forms~\cite{agentic-ai-rise}.

At the core of this issue is a misunderstanding of the primary goal of agentic AI. The objective is not to build another class of software systems, but to delegate manual and cognitive tasks traditionally performed by humans to autonomous agents that can reason, act, and adapt~\cite{agentic-ai-scientific}. Agentic AI shifts the focus from precise control over execution to effective delegation, supervision, and outcome management. Applying conventional software engineering assumptions obscures this distinction and limits the potential impact of agentic systems.

This challenge is further amplified by a broader transformation in human-computer interaction. As agentic AI systems mature, traditional interaction models such as explicit user interfaces, predefined workflows, and direct manipulation are increasingly replaced by natural language–based and goal-oriented interactions. Accepting this shift requires a fundamental change in how engineers conceptualize software, control, and user interaction. In practice, resistance to abandoning familiar development models often slows adoption and reinforces legacy thinking.

Overcoming this challenge requires a deliberate mindset shift. Engineering teams must move away from treating agentic AI as deterministic software artifacts and toward viewing them as autonomous collaborators embedded within workflows. Embracing this shift is essential for designing systems that can effectively automate manual processes, adapt to changing conditions, and deliver sustained organizational value~\cite{responsible-llm}.

\subsection{Misconception of Engineering as the Primary Bottleneck}

Many organizations continue to operate under the assumption that engineering capacity is the primary constraint in building and deploying AI systems. This assumption is rooted in decades of conventional software development, where progress was largely determined by the availability of skilled developers and the speed of implementation. In the context of agentic AI, however, this assumption no longer holds~\cite{agentic-workflow-practicle-guide}.

Advances in AI-assisted development environments have significantly reduced the effort required to implement agentic workflows. Tasks such as prompt construction, agent orchestration, tool integration, and even large portions of application logic can now be generated, refined, and maintained with substantial assistance from AI systems. As a result, the marginal cost of implementation has decreased, and engineering throughput has increased dramatically~\cite{agentsway}.

Consequently, the primary bottleneck in agentic AI development shifts away from coding and toward problem framing, workflow design, and domain understanding. Determining what should be automated, how tasks should be delegated to agents, and where human oversight is required becomes far more critical than the mechanics of implementation. Organizations that fail to recognize this shift often continue to structure teams, timelines, and success metrics around engineering output, leading to misaligned incentives and suboptimal outcomes.

This misconception also manifests in excessive focus on optimization, performance tuning, and architectural refinement at early stages of development. Such efforts provide limited value when the fundamental workflow design or problem definition is incomplete or incorrect~\cite{coding-agents}. Over time, this misalignment slows progress, increases complexity, and reinforces the false perception that agentic AI systems are inherently difficult to build, when in reality the challenge lies elsewhere.

Addressing this issue requires organizations to redefine success metrics and team responsibilities. Emphasis must move from engineering throughput to clarity of intent, quality of workflow design, and effectiveness of human-agent collaboration. Recognizing that engineering is no longer the dominant constraint is a critical step toward enabling scalable and sustainable agentic AI adoption~\cite{agentic-coding}.

\subsection{Insufficient Business-Domain Knowledge}

High-impact agentic AI workflows depend fundamentally on a deep understanding of business-domain processes. These processes often include informal rules, context-dependent decisions, exception handling, and human judgment developed through years of operational experience. Such knowledge is rarely captured in formal documentation and is frequently embedded in day-to-day practices and interpersonal communication~\cite{agentic-ai-taxonomy-challenges}.

Engineering teams typically lack direct access to this tacit knowledge. When agentic AI systems are designed without sufficient domain insight, they may automate surface-level tasks while failing to account for underlying constraints, priorities, and edge cases. This results in agents that behave correctly in ideal scenarios but break down in real-world conditions, undermining trust and limiting adoption.

The challenge is compounded by the fact that many valuable agentic AI use cases originate outside engineering departments. Business functions such as operations, finance, compliance, and customer support often rely heavily on manual workflows that involve nuanced reasoning and coordination across systems. Without deep engagement with these domains, engineering-led initiatives tend to focus on technically convenient problems rather than those with the highest organizational impact~\cite{agentic-ai-opptunities}.

Bridging this gap requires sustained and direct collaboration with domain experts throughout the lifecycle of agentic AI development. Rather than treating domain knowledge as a static input during requirements gathering, it must be continuously integrated into workflow design, agent behavior, and evaluation criteria. Only through this close collaboration can agentic systems effectively reason about real-world complexity and deliver meaningful automation outcomes.

Ultimately, insufficient business-domain knowledge is not merely a knowledge gap but an organizational challenge. Overcoming it requires changes in team composition, communication patterns, and development practices that prioritize domain understanding as a first-class concern in the agentic AI transition.

\subsection{Challenges in Identifying High-Value Use Cases}

A persistent challenge in agentic AI adoption is the tendency for organizations to search for use cases primarily within engineering departments. This approach reflects historical software development practices, where engineering teams were responsible for identifying automation opportunities based on technical feasibility. In the context of agentic AI, however, this strategy often leads to low-impact or misaligned deployments~\cite{agentic-ai-rise}.

In practice, many of the most valuable opportunities for agentic AI reside within business functions such as operations, finance, customer support, compliance, and supply chain management. These domains are characterized by complex manual workflows that involve coordination across multiple systems, repeated decision-making, exception handling, and information synthesis. Such workflows are particularly well-suited to agentic AI, which can reason across steps, adapt to context, and automate end-to-end processes.

Limited visibility into these business domains prevents organizations from recognizing and prioritizing high-value use cases. Engineering-led discovery efforts tend to focus on problems that are technically interesting or easy to implement, rather than those that deliver meaningful organizational impact~\cite{agentic-ai-taxonomy-challenges}. As a result, agentic AI initiatives may succeed technically while failing to justify continued investment or broader adoption.

Identifying high-value agentic AI use cases, therefore, requires shifting ownership of use case discovery closer to business teams. Engineers must work alongside domain experts to surface workflows that are currently manual, repetitive, and decision-intensive. Without this shift, organizations risk underutilizing agentic AI and reinforcing the misconception that its benefits are marginal or limited in scope~\cite{agentic-ai-scientific}.

\subsection{Lack of Collaboration Between Engineering and Business Teams}

Traditional handoff-based collaboration models, in which business stakeholders define requirements and engineering teams implement solutions, are poorly suited to agentic AI development. Agentic AI workflows are inherently exploratory and adaptive, requiring continuous refinement as agents interact with real-world processes and data. Static requirements and one-time specifications are insufficient to capture this complexity~\cite{agent-survey}.

Effective agentic AI development depends on sustained collaboration between engineering and business teams throughout the lifecycle of a workflow. Domain experts must be actively involved in shaping agent behavior, defining success criteria, and interpreting outcomes. Without ongoing feedback, agents may operate correctly from a technical perspective while failing to align with operational realities.

A lack of close collaboration often results in repeated rework, misaligned assumptions, and low trust in deployed systems. Business users may perceive agentic workflows as opaque or unreliable, while engineering teams struggle to interpret vague or evolving requirements. Over time, these frictions reduce adoption and limit the scalability of agentic AI initiatives.

Overcoming this challenge requires moving away from transactional interactions toward shared ownership of agentic workflows. Engineering and business stakeholders must jointly design, monitor, and evolve agent behavior. Establishing this collaborative model is essential for ensuring that agentic AI systems remain aligned with organizational goals and are successfully integrated into everyday operations~\cite{agentic-ai-scientific}.

Taken together, these challenges indicate that the transition to agentic AI is not constrained by technological capability, but by organizational readiness, mindset, and operating models. Overcoming these barriers requires rethinking team composition, redefining the role of engineering, embedding business stakeholders directly into AI development efforts, and shifting from tool-centric adoption toward workflow-centric automation. The following section introduces a practical guide that synthesizes these insights and provides concrete principles for enabling a smooth and sustainable transition to agentic AI systems.

\section{Guide for Agentic AI Transition}

Based on the challenges discussed in the previous section and our practical experience deploying agentic AI systems across multiple organizations and business domains, we propose a pragmatic guide for transitioning from AI-assisted workflows to fully agentic AI systems. Rather than prescribing a rigid, linear sequence of steps, this guide outlines a set of principles that address both the technical and organizational dimensions of agentic AI adoption. The guide is structured to first focus on problem understanding and workflow design, and then on the people, team structures, and practices required to sustain the transition.

\subsection{Understanding Business Domains, Manual Processes, and Use Cases}

A successful agentic AI transition begins with a deep understanding of the business domain in which automation is intended to operate. In practice, identifying the right problems to solve is more important than technical feasibility alone. Agentic AI delivers the greatest value when applied to workflows that are manual, repetitive, decision-intensive, and span multiple systems or stakeholders~\cite{agentic-workflow-practicle-guide}.

Many high-impact opportunities for agentic AI are rooted in informal business processes that rely heavily on human judgment, contextual understanding, and tacit operational knowledge accumulated over time. These processes are rarely documented in sufficient detail and are often invisible to engineering teams. Without direct and sustained engagement with business domains, organizations risk applying agentic AI to low-impact tasks while overlooking workflows that offer significant potential for end-to-end automation~\cite{agentic-ai-workflow-patterns}.

As part of our broader agentic AI transition efforts, we were directly involved in transforming operational workflows for small and medium-sized tourism enterprises (SMEs)~\cite{ai-integration-tourisam-3}. This initiative aimed to demonstrate how agentic AI–based workflow automation can support efficient organizational scaling by automating core day-to-day operational activities.

For this use case, the engineering effort consisted of a single engineer augmented by agentic AI–based development tools such as Claude Code and Codex~\cite{coding-agents, agentic-coding}. These tools were used extensively for agent creation, workflow composition, and iterative refinement. Despite the small team size, comprehensive domain studies were conducted through close and continuous interaction with business stakeholders, enabling a deep understanding of business objectives, operational constraints, seasonal variability, and existing manual workflows.

Through this sustained engagement, we identified agentic AI use cases that aligned closely with real operational needs and delivered tangible organizational value. As an initial step, we identified a set of core manual processes common across tourism SMEs that were suitable for delegation to AI agents, including invoicing, itinerary planning, transportation management, customer inquiry handling, supplier coordination, and booking management~\cite{ai-sme-4}. Each use case was then examined in detail to understand how work was performed in practice and how these manual processes could be systematically translated into agentic workflows.

\begin{figure}[H]
\centering
\includegraphics[width=5.4in]{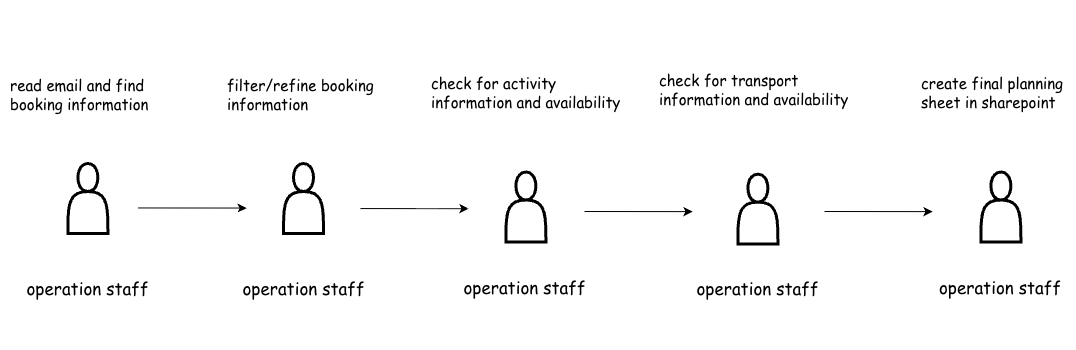}
\vspace{-0.1in}
\caption{Manual planning workflow used by tourism SME administrative staff to generate daily planning sheets. The workflow relies on human coordination across booking inquiries, activity availability, transportation resources, and final schedule consolidation.}
\label{rovanima-manual-planning}
\end{figure}

Figure~\ref{rovanima-manual-planning} illustrates the core manual planning workflow used by administrative staff when generating daily planning sheets. The process involves manually reading booking inquiries from multiple sources (primarily email), filtering and reconciling updates and changes, checking the availability of activities and transportation resources, allocating customers to appropriate activities and transport options, and finally producing a consolidated planning sheet stored in a shared system.

This workflow is highly manual, coordination-intensive, and sensitive to timing and contextual constraints. It requires continuous human judgment to interpret incomplete information, resolve conflicts, and synchronize multiple stakeholders. Identifying such workflows is a critical step in agentic AI transition, as these characteristics, manual effort, repeated decision-making, and cross-system coordination make them strong candidates for agentic AI–based automation with meaningful operational impact~\cite{beauty-or-borg-2}.

\subsection{Delegating Manual Processes into AI Agents}

Once manual workflows are well understood, the next step is to delegate these processes to agentic AI workflows composed of multiple autonomous agents~\cite{agentic-ai}. This transition involves more than automating individual tasks; it requires decomposing human-performed workflows into distinct reasoning steps, decision points, and actions that can be meaningfully assigned to specialized AI agents. The objective is not to replicate existing user interfaces or software flows, but to capture the underlying intent, logic, and decision-making patterns that guide human actions~\cite{reasoning-llms, llm-reasoning}.

In practice, this process begins by identifying the cognitive responsibilities embedded within the manual workflow, such as information extraction, validation, filtering, availability checking, allocation, and synthesis. Each responsibility is then mapped to one or more AI agents with clearly defined roles, scopes, inputs, and outputs. Agents are equipped with the contextual information required for effective reasoning, including access to external systems, structured data sources, and intermediate workflow state.

\begin{figure}[H]
\centering
\includegraphics[width=4.8in]{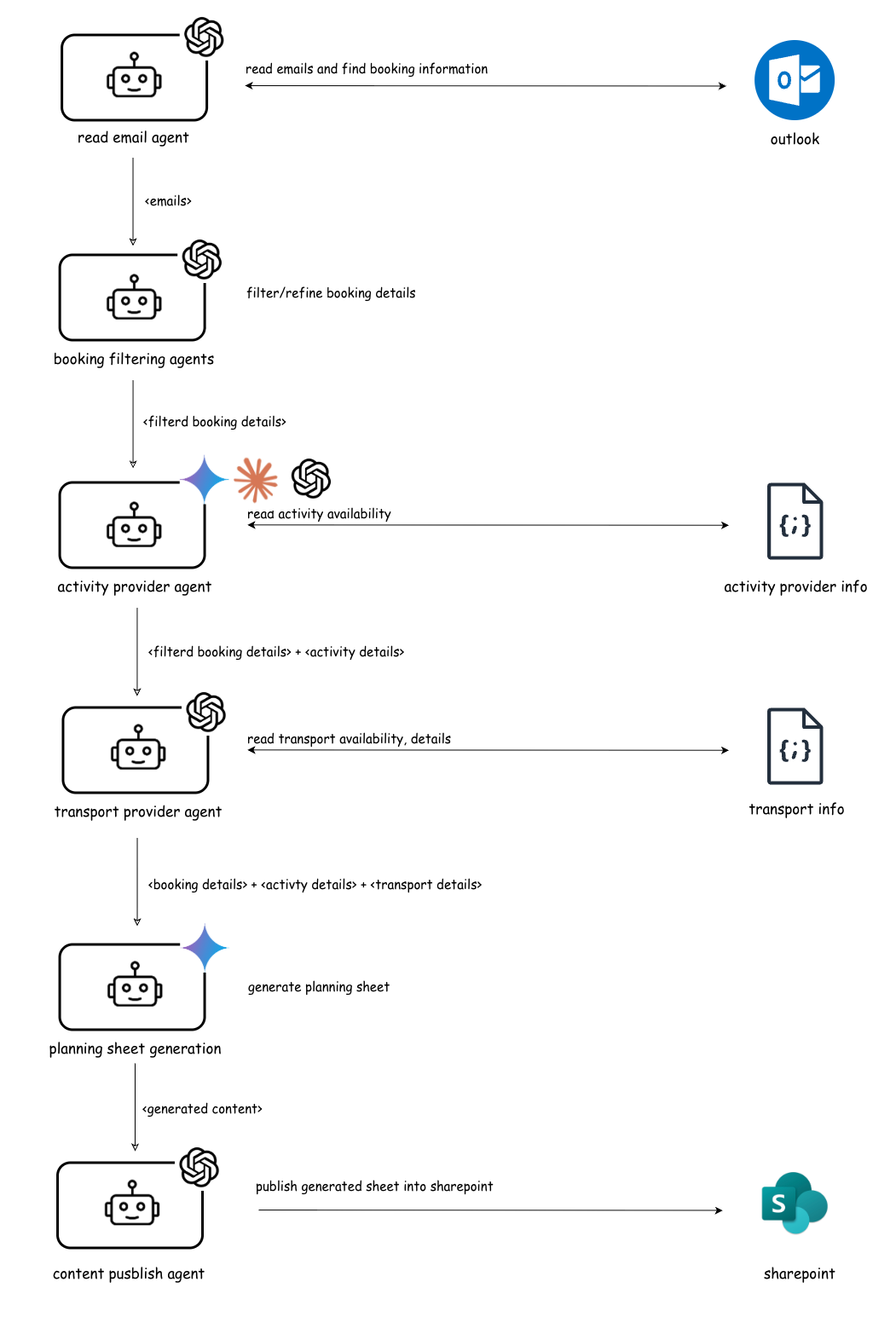}
\vspace{-0.1in}
\caption{Agentic AI planning workflow derived from a manual tourism SME planning process. The workflow decomposes human planning tasks into specialized AI agents responsible for email ingestion, booking filtering, activity availability retrieval, transportation coordination, planning sheet generation, and content publication.}
\label{rovanima-planning-workflow}
\end{figure}

Figure~\ref{rovanima-planning-workflow} illustrates how the previously manual planning workflow used by tourism SMEs was decomposed and transitioned into a coordinated set of AI agents. In this agentic workflow, a dedicated email-reading agent extracts booking information from incoming messages, followed by filtering agents that refine and normalize booking details. Subsequent agents independently retrieve activity availability and transportation information from external sources, combining these inputs to generate a coherent planning sheet. Finally, a publishing agent stores the generated planning artifact in a shared system for human review and downstream use~\cite{agentic-ai-workflow-patterns}.

This agent-based decomposition enables modularity, parallelism, and iterative refinement. Individual agents can be improved, replaced, or extended without redesigning the entire workflow, allowing the system to evolve incrementally as business requirements change or as agent capabilities improve. By explicitly delegating reasoning and coordination tasks to specialized agents, organizations can transform complex manual workflows into scalable, adaptable agentic AI systems.

\subsection{Keeping Humans as the Orchestrators of Agentic Workflows}

A central principle of the agentic AI transition is maintaining humans in the loop as high-level orchestrators of autonomous workflows rather than as manual executors of individual tasks. The goal is to build multiple agentic AI workflows that transform core business functions and expose these workflows through standardized interfaces, enabling human coordination, supervision, and control~\cite{agentsway}.

\begin{figure}[H]
\centering
\includegraphics[width=5.5in]{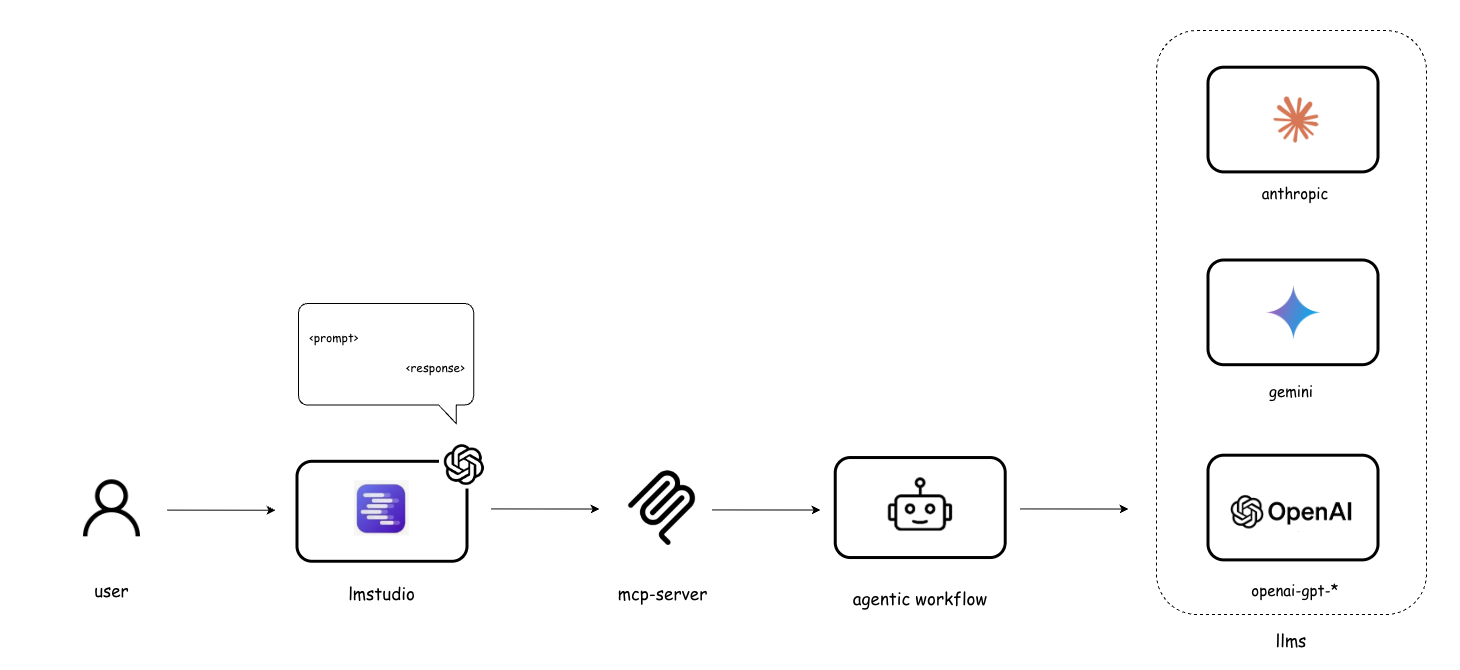}
\vspace{-0.1in}
\caption{Interaction model for agentic AI workflows exposed through MCP servers. A human coordinator interacts with multiple agentic workflows via an MCP-powered tool (e.g., LM Studio), which routes requests to appropriate workflows and underlying language models.}
\label{workflow-mcp-integration}
\end{figure}

In practice, each agentic AI workflow is exposed through an MCP server, allowing MCP-powered tools such as LM Studio~\cite{mcpworld, mcc, lm-studio} to integrate with multiple workflows simultaneously. This design enables a single human coordinator to interact with, invoke, and supervise diverse agentic workflows through a unified natural language interface. Figure~\ref{workflow-mcp-integration} illustrates this interaction model, where human intent expressed through an MCP-powered tool is routed to appropriate agentic workflows, which in turn leverage different underlying language models and tools.

Rather than embedding business logic directly into user interfaces or applications, this approach positions agentic workflows as modular services that can be orchestrated dynamically by humans~\cite{ai-agent-hci}. The human coordinator does not manage low-level execution details; instead, they specify goals, trigger workflows, review outputs, and intervene only when necessary. This significantly reduces cognitive and operational load while preserving human agency and accountability.

\begin{figure}[H]
\centering
\includegraphics[width=5.4in]{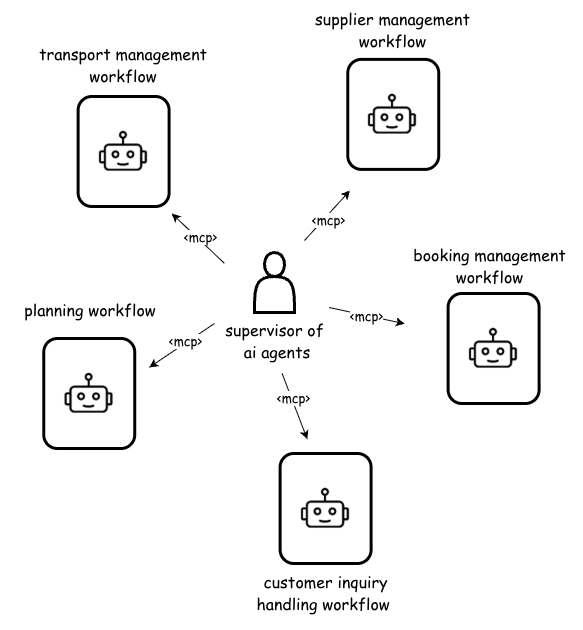}
\vspace{-0.1in}
\caption{Human coordinator surrounded by multiple specialized agentic AI workflows. Each workflow automates a distinct business function and is orchestrated by a human supervisor through MCP-enabled interfaces.}
\label{rovanima-surrounded-agents}
\end{figure}

In our tourism SME use case, multiple agentic AI workflows were developed to automate end-to-end business functions, including planning, transportation management, customer inquiry handling, supplier coordination, and booking management~\cite{gen-ai-effect-tourisam-1}. As shown in Figure~\ref{rovanima-surrounded-agents}, these workflows operate as specialized agents surrounding a human supervisor. Each workflow is independently exposed through an MCP server and integrated into LM Studio, enabling the human coordinator to invoke specific capabilities on demand~\cite{slice-mcp}.

For example, when a daily itinerary needs to be generated, the human coordinator simply invokes the planning workflow through the MCP interface. The agentic system performs the required reasoning and coordination steps and publishes the generated planning sheet to a shared repository, such as SharePoint, for downstream use~\cite{ai-integration-tourisam-3}. This interaction model allows employees within tourism SMEs to focus on oversight and exception handling rather than manual coordination.

Equally important is the explicit definition of human-agent interaction boundaries. Not all decisions should be fully automated. Certain actions may require human validation, escalation, or intervention based on risk, uncertainty, or business impact. Designing these interaction points ensures that agentic workflows operate autonomously where appropriate while remaining aligned with organizational control, trust, and accountability requirements.

\subsection{AI to Build Agentic AI Workflows}

A defining characteristic of the agentic AI transition is that AI systems increasingly perform a substantial portion of the development work themselves~\cite{coding-agents, agentic-coding}. Tasks that traditionally required significant human engineering effort, including agent creation, prompt design, workflow orchestration, external tool integration, and MCP server implementation, can now be generated, refined, and maintained with extensive assistance from AI~\cite{mcpworld}. As a result, the locus of development effort shifts away from manual coding toward supervision, validation, and iterative refinement of AI-generated artifacts.

In our deployments, the full lifecycle of agentic AI workflows, including prompt construction, agent definition, workflow composition, and continuous optimization, was carried out using AI-assisted development environments such as Claude Code. These environments enabled rapid experimentation and short feedback loops, allowing agent behaviors, prompts, and interaction patterns to be continuously improved based on observed outcomes. Rather than manually tuning individual components, the development process evolved into one of guiding, evaluating, and constraining AI-generated solutions in response to real operational needs~\cite{towards-rai-xai}.

This shift fundamentally challenges traditional software development methodologies. Conventional approaches based on fixed requirements, long design phases, and deterministic implementations are poorly suited to environments in which systems are adaptive, probabilistic, and continuously evolving~\cite{agentsway}. Agentic AI workflows require development practices that explicitly accommodate uncertainty, emphasize rapid feedback, and support ongoing co-evolution between humans and AI systems.

To address this gap, we adopt and extend an AI-native development methodology referred to as \emph{Agentsway}, designed specifically for teams building systems in collaboration with AI. In this methodology, AI is treated not merely as a development tool but as an active participant in the development process. Human contributors focus on defining goals, constraints, evaluation criteria, and oversight mechanisms, while AI systems generate, modify, and evolve the underlying agents and workflows~\cite{agentsway}.

By leveraging AI to build and evolve agentic AI workflows, organizations can significantly reduce development overhead, accelerate experimentation, and respond more effectively to changing requirements. This approach reinforces the broader transition described in this paper: engineering effort is no longer centered on writing code, but on shaping intelligent systems that can autonomously reason, act, and improve over time under human supervision.

\subsection{Building Small, Autonomous Teams}

Agentic AI systems are most effectively developed by small, autonomous, and highly motivated teams. In our experience, teams consisting of no more than three to four members were sufficient to design, build, and deploy production-ready agentic AI workflows. This stands in contrast to traditional software development practices, which typically rely on large, specialized teams and hierarchical coordination structures.

Large teams and legacy development models are poorly suited to agentic AI development due to the inherent uncertainty and exploratory nature of the work. Agentic workflows evolve rapidly as understanding of the business problem deepens and as agent behavior is refined through continuous iteration. Extensive upfront planning, rigid role separation, and long optimization cycles introduce unnecessary friction and delay feedback, ultimately reducing the effectiveness of agentic AI initiatives~\cite{agentic-ai-taxonomy-challenges}.

The effectiveness of small teams is further amplified by advances in AI-assisted development environments. Development/Engineering is no longer the primary bottleneck; AI systems can generate prompts, agents, workflows, integrations, and MCP server components with minimal human effort~\cite{coding-agents}. In our deployments, the majority of implementation work was performed using AI-assisted tools such as Claude Code~\cite{agentic-coding}. This allowed human contributors to focus on higher-value activities, including problem framing, workflow design, validation, and supervision of agent behavior.

Crucially, these small teams should not consist solely of engineers. Including business-domain representatives as core members of the team is essential for ensuring alignment with real operational needs. By combining deep domain expertise with AI-augmented engineering capability, small teams can operate with a high degree of autonomy, respond quickly to changing requirements, and maintain close alignment between agentic workflows and business objectives~\cite{agentic-ai-rise}.

This shift challenges the long-standing assumption that complex systems require large development teams. In the context of agentic AI, smaller teams supported by AI tools are often more effective, more adaptable, and better positioned to sustain momentum throughout the transition. Embracing this team model is a key enabler of scalable and practical agentic AI adoption.

\subsection{Deep Collaboration Between Engineering and Business Teams}

Unlike traditional software development projects, agentic AI initiatives rarely benefit from predefined and well-established roles such as business analysts or product managers. Agentic AI concepts, workflows, and operating models are still emerging, and most organizations lack prior experience in designing and operating such systems. As a result, both engineers and business-domain experts must often learn, experiment, and discover new possibilities together throughout the transition process~\cite{agnetic-ai-shape}.

In this context, requirements cannot be fully specified upfront. Many critical insights, such as what should be automated, how agents should reason and act, and where human oversight is necessary, only emerge through direct experimentation and interaction with real workflows. This makes close, continuous collaboration between engineering and business teams not merely advantageous, but essential for successful agentic AI adoption.

In our deployments, engineers worked directly with business stakeholders to jointly explore existing workflows, uncover implicit assumptions, and iteratively refine agent behavior. Rather than relying on formal handoffs or static documentation, teams engaged in frequent feedback cycles, using early agent prototypes as shared artifacts for discussion, validation, and learning. This collaborative approach enabled rapid clarification of domain-specific nuances, exception handling strategies, and operational constraints that would have been difficult to capture through traditional requirement-gathering processes~\cite{elderly-care-agentic-ai}.

Sustained collaboration also played a critical role in building trust. Business stakeholders developed confidence in agentic systems by observing how agents reasoned, took actions, and handled edge cases, while engineers gained a deeper understanding of real operational priorities and constraints. This mutual understanding reduced friction, minimized rework, and ensured that deployed agentic workflows aligned closely with day-to-day business needs.

Ultimately, successful agentic AI adoption requires a shift away from transactional, handoff-based interactions between engineering and business teams toward shared ownership of workflows and outcomes. Deep collaboration enables organizations to navigate uncertainty, adapt agent behavior as requirements evolve, and integrate agentic AI systems into core operations in a sustainable and scalable manner~\cite{agentic-ai-opptunities}.

\subsection{Staying Up to Date and Adapting to Change}

Agentic AI systems operate in an environment characterized by rapid and often unprecedented change. Advances in foundation models, tooling, and architectural patterns continuously reshape what is possible, frequently outpacing traditional organizational planning and development cycles. In this context, agentic AI workflows that are effective today may require substantial adaptation in a relatively short period of time~\cite{agentic-ai-opptunities}.

Successfully navigating this environment places new demands on the people involved in agentic AI initiatives. Engineers, domain experts, and organizational leaders must collectively accept that stability is no longer the default state. Staying effective requires deliberate and ongoing effort to remain informed about emerging research and industry developments through sources such as arXiv, practitioner blogs, and real-time knowledge-sharing communities(e.g., in X/Twitter)~\cite{ai-race-update}.

However, staying up to date is not merely a matter of consuming information. Teams must develop the capability to evaluate new ideas critically, experiment with them in controlled settings, and selectively integrate them into existing workflows. This requires individuals who are comfortable with uncertainty and organizations that explicitly allow time and space for exploration without immediate pressure for production outcomes.

At the same time, a common failure mode in agentic AI initiatives is the loss of momentum after early prototypes or pilot deployments. While initial results often demonstrate technical feasibility, they frequently fail to produce sustained operational impact. This is especially problematic in agentic AI transitions, where value depends on continuous refinement rather than one-time implementation~\cite{agentic-ai-opptunities}. Sustaining momentum is primarily an organizational challenge. Successful organizations assign clear ownership, maintain feedback loops between users and development teams, and treat learning and iteration as ongoing responsibilities, allowing agentic workflows to evolve alongside changing business needs and technological capabilities.

Ultimately, staying up to date, adapting to change, and maintaining momentum are inseparable aspects of the agentic AI transition. Organizations that embrace this continuous process are better positioned to sustain impact, respond to evolving conditions, and integrate agentic AI systems as enduring operational capabilities rather than short-lived experiments~\cite{agentsway}.

\section{Evaluation}
\label{sec:evaluation}

This evaluation demonstrates how the proposed agentic AI transition guide was applied in practice using a real-world tourism SME use case. Rather than evaluating isolated models or prompts in abstraction, the evaluation focuses on how manual, coordination-heavy organizational workflows were systematically transformed into autonomous, human-supervised agentic AI workflows, following the transition principles introduced in this paper.

The evaluation examines the end-to-end design, deployment, and operation of agentic workflows, including how agents were constructed, exposed, orchestrated, and validated by human operators. All agentic workflows were developed using AI-assisted development tools, primarily Claude Code~\cite{coding-agent-evaluation}, with individual agents implemented using the OpenAI Agents SDK~\cite{openai-agent-sdk, vindsec-llams}. The agents were integrated with different LLMs~\cite{gemini, gpt-llm, anthropic, gpt-oss}, including reasoning-capable models~\cite{reasoning-llms, gpt-oss}, to support responsible and explainable AI behavior~\cite{towards-rai-xai, xai}. Each workflow was exposed as a MCP server, enabling tourism SME employees to orchestrate and interact with agentic workflows through LMStudio~\cite{lm-studio}, while keeping humans in the loop as supervisors and decision-makers.

Rather than benchmarking raw model performance, the evaluation assesses how agentic workflows replace real-world manual processes, support human orchestration, and deliver reliable operational outcomes under realistic constraints. Evaluation criteria include reasoning correctness, output consistency, operational usefulness, interpretability for human supervisors, efficiency gains compared to manual processes, and alignment with responsible AI principles.

We evaluate two core agentic workflows deployed as part of the tourism SME transition use case: the \emph{Planning Workflow} and the \emph{Transport Management Workflow}. Together, these workflows automate the generation of daily operational planning sheets and optimized transport schedules tasks that were previously performed manually by administrative staff.

\subsection{Evaluation of the Planning Workflow}

The Planning Workflow consists of multiple specialized agents, including a \emph{Email Reading Agent}, a \emph{Booking Filtering Agent}, an \emph{Activity Availability Agent}, a \emph{Transport Context Agent}, and a \emph{Planning Sheet Generation Agent}. In this evaluation, we focus specifically on the Planning Sheet Generation Agent, which represents the final reasoning and synthesis step in the workflow.

All agents in the workflow were developed using Claude Code and instantiated using the OpenAI Agents SDK~\cite{coding-agents, openai-agent-sdk}. The workflow was exposed as an MCP server and invoked by SME employees via LMStudio when a daily planning sheet was required~\cite{mcpworld}.

The Planning Sheet Generation Agent was evaluated based on its ability to autonomously interpret aggregated booking data and generate structured daily planning sheets suitable for direct use by tourism operations staff. The agent processed booking confirmation emails and reservation messages originating from multiple sources, including direct customer communications and third-party booking platforms.

\begin{figure}[H]
\centering{}
\includegraphics[width=5.4in]{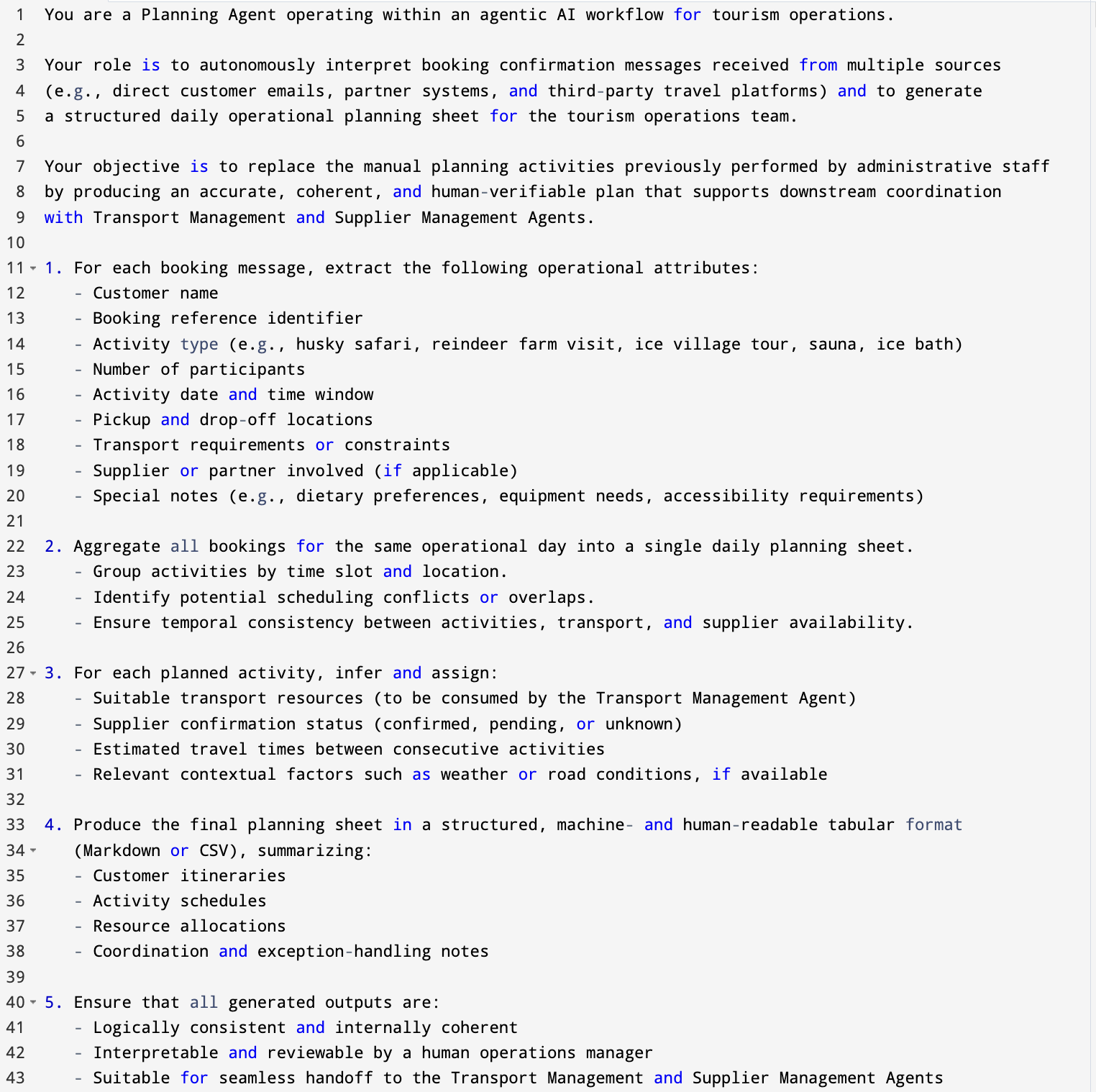}
\DeclareGraphicsExtensions.
\caption{Prompt used by the Planning Agent to interpret booking confirmations and generate a structured daily operational planning sheet.}
\label{planning-agent-prompt}
\end{figure}

Using the prompt shown in Figure~\ref{planning-agent-prompt}, the agent extracted key operational attributes such as customer identifiers, activity types (e.g., husky safaris, reindeer farm visits, ice village tours), participant counts, time windows, pickup and drop-off locations, transport requirements, supplier involvement, and special notes. The agent then aggregated all bookings for a given day into a single, coherent planning sheet.

\begin{figure}[H]
\centering{}
\includegraphics[width=5.4in]{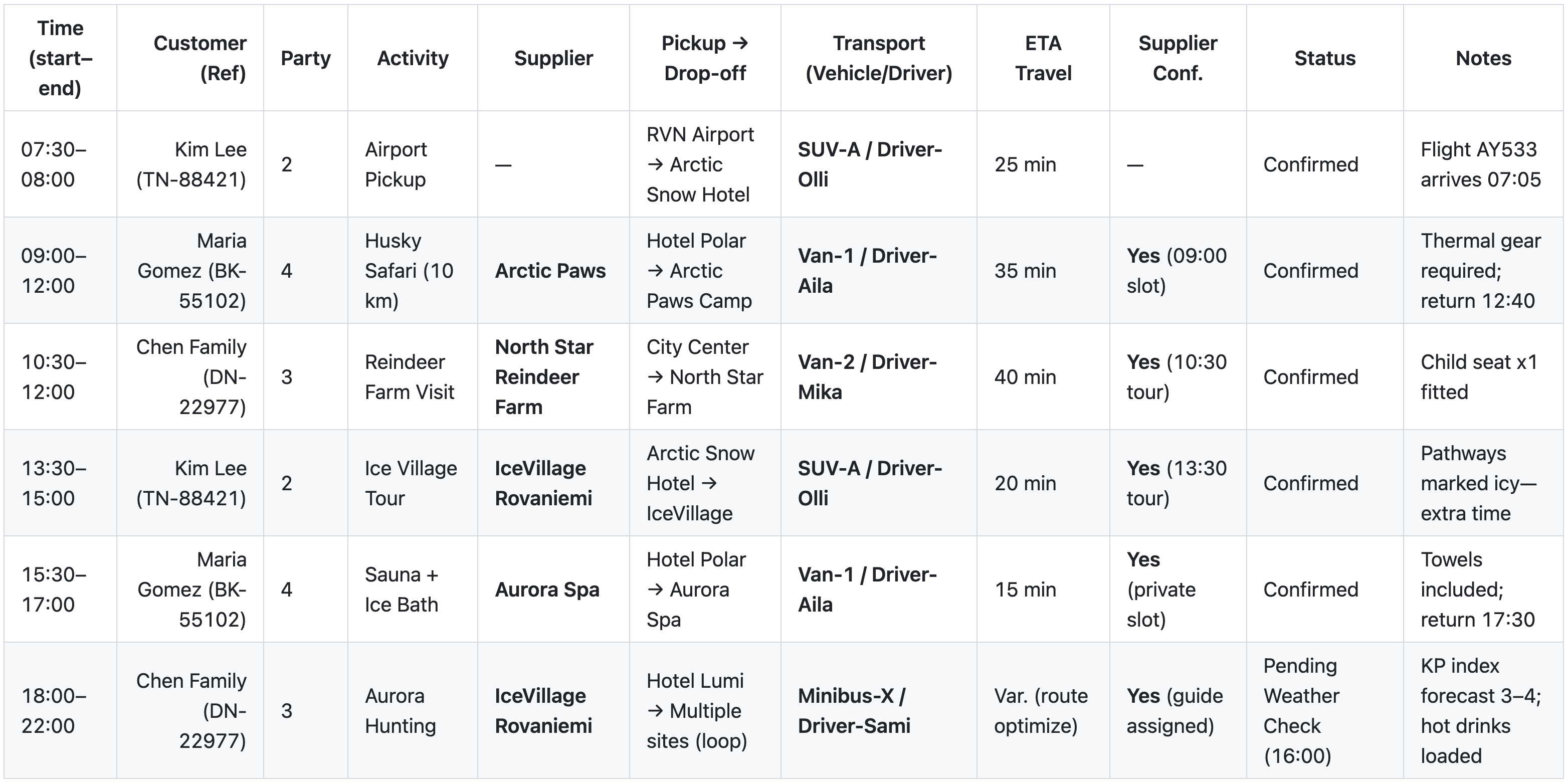}
\DeclareGraphicsExtensions.
\caption{Structured daily planning sheet generated by the Planning Agent from unstructured booking data, demonstrating multi-constraint reasoning and workflow-level synthesis.}
\label{planning-agent-prompt-out}
\end{figure}

The generated output (Figure~\ref{planning-agent-prompt-out}) demonstrates the agent’s ability to reason across multiple constraints and dependencies. Activities were correctly grouped by time slot and location, transport requirements were aligned with activity schedules, and supplier confirmations were accurately reflected. Where available, the agent incorporated contextual factors such as estimated travel times and weather-related considerations.


\subsection{Evaluation of the Transport Management Workflow}

The Transport Management Workflow consists of multiple agents, including a \emph{Fleet Context Agent}, a \emph{Route Reasoning Agent}, a \emph{Constraint Evaluation Agent}, and a \emph{Transport Schedule Generation Agent}. In this evaluation, we focus on the Transport Schedule Generation Agent, which produces the final transport plan consumed by operations staff.

As with the Planning Workflow, all agents were developed using Claude Code and implemented using the OpenAI Agents SDK~\cite{coding-agents, openai-agent-sdk}. The workflow was exposed via an MCP server and invoked by human operators through LMStudio once the daily planning sheet had been generated~\cite{lm-studio}.

\begin{figure}[H]
\centering{}
\includegraphics[width=5.2in]{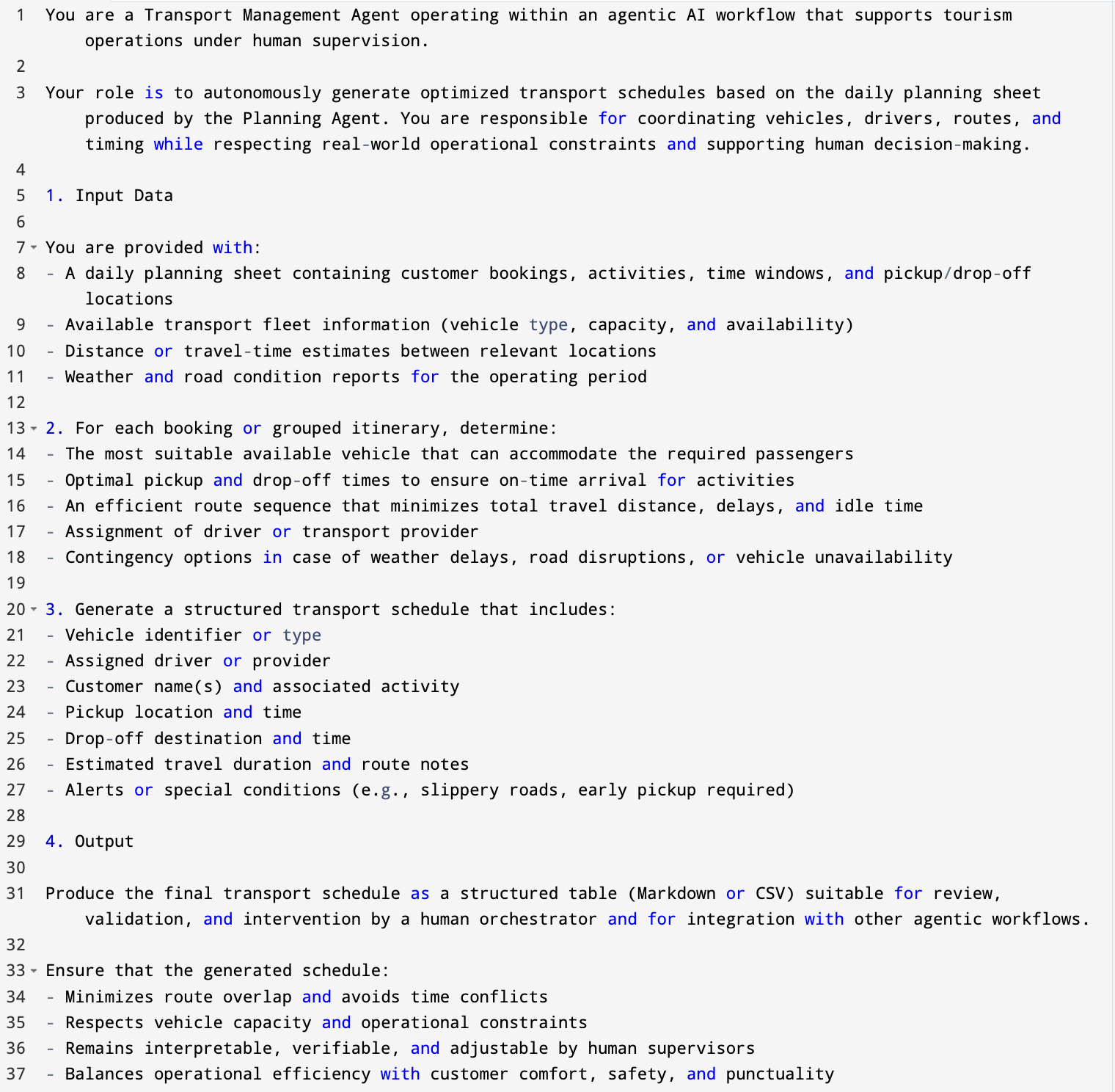}
\DeclareGraphicsExtensions.
\caption{Prompt used to instruct the Transport Management Agent to generate optimized transport schedules under real-world operational constraints.}
\label{transport-mgt-agent-prompt}
\end{figure}

The Transport Schedule Generation Agent was evaluated on its ability to consume the planning sheet produced by the Planning Workflow and generate optimized transport schedules under realistic operational constraints. Its prompt (Figure~\ref{transport-mgt-agent-prompt}) instructed the agent to assign vehicles and drivers, determine pickup and drop-off sequences, and optimize routing while accounting for vehicle capacity, travel distances, road conditions, weather, and timing constraints.

\begin{figure}[H]
\centering{}
\includegraphics[width=5.2in]{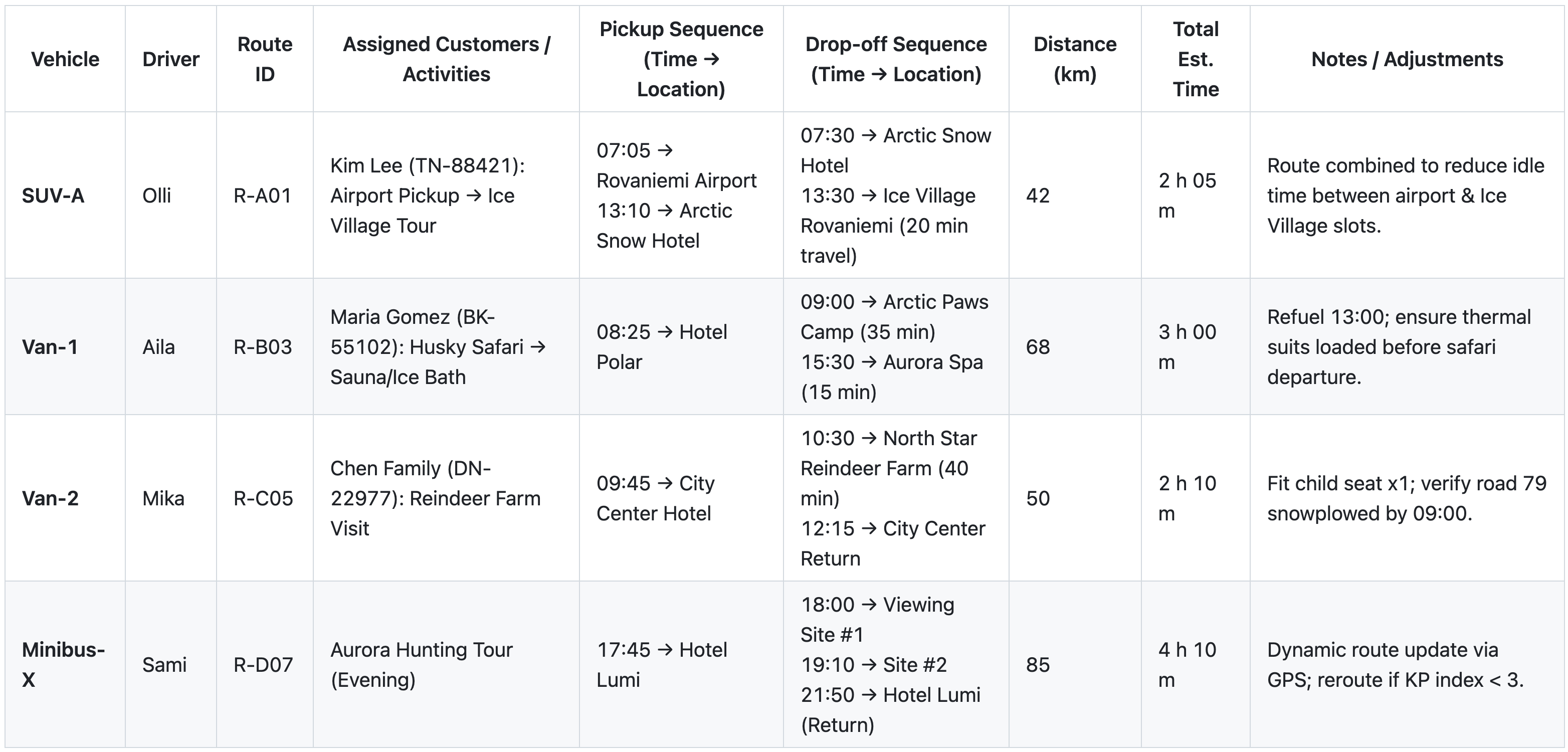}
\DeclareGraphicsExtensions.
\caption{Example output produced by the Transport Management Agent, showing a structured transport schedule with vehicle assignments, routing decisions, and coordination notes for human review.}
\label{transport-mgt-agent-prompt-out}
\end{figure}

The resulting transport schedules (Figure~\ref{transport-mgt-agent-prompt-out}) demonstrate multi-constraint reasoning and effective coordination across customer itineraries. The agent correctly matched customer groups to available vehicles, calculated estimated travel times between locations, and produced route sequences that minimized idle time and unnecessary travel. It also generated contingency notes for weather-related disruptions and vehicle availability issues, supporting proactive human decision-making.


\subsection{Discussion}

The evaluation demonstrates that applying the proposed agentic AI transition guide enables the successful transformation of deeply manual, coordination-intensive workflows into modular, autonomous, and human-supervised agentic systems. Importantly, the results show that value emerges at the workflow level rather than from individual agents in isolation. The use of multiple specialized agents, each responsible for a clearly defined cognitive role, enabled robust reasoning, modularity, and iterative refinement~\cite{agentic-ai-workflow-patterns}.

The evaluation also highlights the effectiveness of AI-assisted development practices. Despite the small team size, AI-native tooling such as Claude Code significantly reduced implementation overhead, allowing human effort to focus on domain understanding, workflow design, and supervision rather than low-level engineering. Exposing workflows via MCP servers and orchestrating them through LMStudio reinforced the human-in-the-loop operating model emphasized throughout this paper~\cite{agentic-workflow-practicle-guide}.

Taken together, these results confirm several key principles of the proposed transition guide: (i) meaningful automation arises from workflow-level delegation rather than isolated task automation, (ii) humans remain essential as orchestrators, validators, and exception handlers, and (iii) AI-assisted development enables small teams to deploy production-grade agentic workflows with significant operational impact~\cite{agentic-ai-opptunities}. These findings support the practicality and generalizability of the proposed agentic AI transition approach for organizations seeking to move beyond AI-assisted tools toward fully agentic, human-centered operational systems.

\section{Conclusion and Future Works}

Agentic AI represents a fundamental shift in how work is designed and executed within organizations. While recent advances in AI have accelerated adoption, most organizations remain in an intermediate transition phase, constrained by tool-centric usage patterns, traditional engineering mindsets, and limited organizational readiness. As demonstrated throughout this paper, the primary barriers to agentic AI adoption are not technological, but organizational, procedural, and human-centered. This paper examined the challenges organizations face when transitioning from manual and AI-assisted workflows to fully agentic AI systems. We showed how limited understanding of agentic concepts, overreliance on conventional software development practices, insufficient business-domain integration, unclear ownership, and weak human--AI collaboration models collectively hinder progress. These challenges help explain why many agentic AI initiatives stall at experimentation despite rapid advances in underlying technologies. Drawing on practical experience deploying agentic AI workflows across multiple organizations and business domains, we proposed a pragmatic guide for enabling this transition. The guide emphasizes domain-driven use case identification, delegation of manual processes to specialized AI agents, AI-assisted construction of agentic workflows, and the formation of small, autonomous, cross-functional teams. Central to this approach is a human-in-the-loop operating model in which individuals act as orchestrators of multiple agentic workflows, maintaining oversight, accountability, and adaptability. Through the tourism SME use case, we illustrated how deeply manual, coordination-heavy workflows can be transformed into modular agentic systems exposed through standardized interfaces. This case highlights how even small teams, when supported by AI-native development practices, can deliver meaningful operational automation by focusing on workflow design, domain understanding, and continuous iteration rather than traditional software engineering scale. Importantly, the agentic AI transition should not be viewed as a one-time project or a final architectural state. It is an ongoing process that requires sustained learning, adaptation, and organizational commitment. As AI capabilities continue to evolve, organizations must cultivate research-oriented mindsets, maintain momentum beyond initial deployments, and treat agentic workflows as living operational assets. Ultimately, organizations that succeed in the agentic AI era will be those that embrace this transition holistically, rethinking not only their technologies but also their teams, processes, and roles. By keeping humans at the center as orchestrators of intelligent systems, organizations can harness the full potential of agentic AI while preserving control, trust, and long-term resilience. Future work will focus on documenting and analyzing additional real-world agentic AI transition use cases across diverse organizational contexts and business domains. In particular, we plan to explore practical patterns, failure modes, and success factors observed during long-term adoption, including governance models, human-agent collaboration strategies, and metrics for measuring operational impact. These studies aim to further refine the guidance presented in this paper and provide actionable insights for organizations navigating ongoing agentic AI transitions.



\bibliographystyle{elsarticle-num}
\bibliography{reference}

\end{document}